\documentclass{aa}
\usepackage{psfig}

\def \rsun {\ifmmode$R$_{\odot}\else R$_{\odot}$\fi}

\def \hcm {\hbox {\ifmmode $ H atom cm$^{-2}\else H atom cm$^{-2}$\fi}}
\def \src {XTE J0421+560}
\def\approxgt{\mathrel{\hbox{\rlap{\lower.55ex \hbox {$\sim$}}
        \kern-.3em \raise.4ex \hbox{$>$}}}}
\def\approxlt{\mathrel{\hbox{\rlap{\lower.55ex \hbox {$\sim$}}
        \kern-.3em \raise.4ex \hbox{$<$}}}}

\newcommand {\asca} {ASCA}

\newcommand {\sax} {BeppoSAX}

\psfigurepath{/usr4/users/aorr/mcg-6-30-15}

\begin{document}

\thesaurus{ (08.02.5; 08.05.2; 08.09.2; 08.14.2; 13.25.5)}

\title{The extraordinary X-ray spectrum of \src}

\author{A.~Orr\inst{1} 
\and A.N.~Parmar\inst{1}
\and M.~Orlandini\inst{2}
\and F. Frontera\inst{2,3}
\and D.~Dal~Fiume\inst{2}
\and A. Segreto\inst{4}
\and A. Santangelo\inst{4}
\and M. Tavani\inst{5,6}
}
\institute
{Astrophysics Division, Space Science Department of ESA, 
ESTEC, 2200 AG Noordwijk, The Netherlands
\and
Istituto Tecnologie e Studio Radiazioni Extraterrestri, CNR, Via Gobetti 101, 
40129 Bologna, Italy
\and
Dipartimento di Fisica, Universit\`a di Ferrara, Via Paradiso 11, 44100,
Ferrara, Italy
\and
Istituto di Fisica Cosmica ed Applicazioni all'Informatica, CNR, Via 
U. La Malfa 153, 90146 Palermo, Italy
\and 
Columbia Astrophysics Laboratory, Columbia University, 538 West 120th Street,
New York, NY 10027, USA
\and
Istituto di Fisica Cosmica e Tecnologie Relative, CNR, Via Bassini 15, 
20133 Milano, Italy
}

\date{Received ; accepted}

\offprints{aparmar@astro.estec.esa.nl}

\maketitle
\markboth{A. Orr et al.: \sax\ spectroscopy of \src}{A. Orr et al.:
\sax\ spectroscopy of \src}

\begin{abstract}
We report results of two \sax\ observations of the transient
X-ray source \src\ during the outburst that started in March 1998.
The source exhibits radio jets and coincides with the binary system CI~Cam.
The 0.1--50~keV spectrum is unlike those of other X-ray
transients, and cannot be fit with any simple model.
The spectra can be represented by an absorbed
two component bremsstrahlung model with narrow Gaussian emission
features identified with O, Ne/Fe-L, Si, S, Ca and Fe-K.
During the second observation (TOO2) the energies of
the O and the Ne/Fe-L features decreased smoothly by $\sim$9\% 
over an interval of 30~hrs. 
No significant energy shift of the other lines
is detected with e.g, a 90\% confidence upper limit to any Fe-K line shift of
3.5\%. 
No moving lines were detected during the first
observation (TOO1) with e.g., an upper limit of
$<$1.4\% to any shift of the Fe-K line. 
The low-energy absorption decreased by a factor $\ge$1.8
from $\sim$$6 \times 10^{21}$~atom~cm$^{-2}$ between TOO1 and TOO2.
We propose that the time variable emission lines  
arise in precessing relativistic jets, while
the stationary lines originate in circumstellar material.
 
\end{abstract}

\keywords{stars: binaries --- stars: emission-line, Be 
-- stars: individual (\src) -- stars: novae -- 
X-rays: stars}

\section{Introduction}
\label{sec:introduction}

Relativistic galactic jet sources are usually seen in X-ray binaries, where 
a black hole or neutron star is accreting from a ``normal'' star.
Recently a new galactic jet source,
\src, was discovered by the Rossi-XTE satellite as a bright and rapidly 
rising X-ray transient on 1998 March 31 (\cite{smith98}).
The source reached a peak intensity of $\sim$2 Crabs on April~1, before rapidly
decaying. Radio observations quickly allowed the identification of the 
optical counterpart with the binary system CI Cam (= MWC 84). 
\cite{miroshnichenko95} and \cite{bergner95} model 
CI Cam as a K0~{\sc ii}--B0~{\sc v} system embedded in 
a hot circumstellar dust shell.  
\cite{chkhikvadze70} estimates the interstellar extinction (1.5 mag.) and 
the distance (1~kpc) of CI Cam. Optical spectra before and after the outburst 
exhibit strong Balmer and He~{\sc i} emission lines (\cite{merrill33}).  
He~{\sc ii} lines appeared after the outburst 
(\cite{wagner98}).
Within a week of the X-ray outburst, extended radio emission appeared
in the form of an S-shaped jet.
If the X-ray and radio outbursts are assumed contemporaneous, for a 
distance of 1~kpc, the rate of expansion of the radio emission implies a 
tangential velocity of 0.15~c (\cite{hjellming98}).

In this {\it Letter} we describe the \sax\
spectra of \src\ in the days following the 
1998 March 31 outburst. X-ray and optical timing analysis and 
light curves are presented in \cite{frontera98}.

\section{Observations and data reduction}
\label{observations}

Data from the Low-Energy Concentrator Spectrometer (LECS;
0.1--10~keV), Medium-Energy Concentrator
Spectrometer (MECS; 1.3--10~keV)
High Pressure Gas Scintillation Proportional Counter
(HPGSPC; 5--120~keV) and the Phoswich
Detection System (PDS; 15--300~keV) on-board \sax\ (\cite{boella97})
are presented. All these instruments are co-aligned and collectively referred
to as the Narrow Field Instruments, or NFI. 

\begin{table*}
\caption[]{Results of fits to the \sax\ NFI spectra.
Model code: PL = power-law; CO PL = cut-off power-law; BKNPL = broken 
power-law;
DBBPL = disk blackbody + power-law; 2BRMS = double bremsstrahlung; 2MEKAL =
double MEKAL. Except for the 2MEKAL model, emission lines fixed at 
the energies given in Table \ref{fits} are included. 
The cut-off and break energies are listed for the CO PL 
and BKNPL
models, respectively. The metal abundance ``Fe/He'' with respect to solar is
given for the 2MEKAL model. 
$\rm{N_H}$ is in units of $\rm {10^{21}}$ atom $\rm {cm^{-2}}$, kT and E are 
in keV}
\begin{flushleft}
\begin{tabular}{llllrlllr}
\hline\noalign{\smallskip}
 & \multicolumn{4}{c}{TOO1} & \multicolumn{4}{c}{TOO2}\\ 
\noalign{\smallskip\hrule\smallskip}
Model  &  \hfil N$_{\rm {H}}$ \hfil & kT, E$_{\rm co,br}$  
&\hfil $\Gamma$, Fe/He \hfil 
& $\chi^2_{\nu}$/dof &
\hfil N$_{\rm {H}}$ \hfil & kT, E$_{\rm co,br}$  &\hfil $\Gamma$, Fe/He \hfil 
& $\chi^2_{\nu}$/dof\\
\noalign{\smallskip\hrule\smallskip}
%
PL & $16.3 \pm 0.4$ & \dots & $2.58 \pm 0.01$ 
& 5.3/719 &
 $2.0 \pm 0.1$ & \dots & $3.08 \pm 0.02$ 
& 5.19/495\\
%
CO PL & $6.0 \pm 0.2$ & $7.6 \pm 0.3$  &$1.53 \pm 0.03$   
& 1.23/718 &
 $2.0 \pm 0.1$ & $>$135 & $3.07 \pm 0.02$
& 5.21/493\\
%
BKNPL  & $16.5 \pm 0.4$ & $1.00 \pm 0.03$ 
& $11.3\, \pm\, ^{1.5}_{1.2}$ & 5.2/717 &
$3.2 \pm 0.1$ & $1.84 \pm 0.04$
&  $5.54 \pm 0.09$ & 1.64/492 \\
 & \dots& \dots& $2.58 \pm 0.01$ & & \dots& \dots& $2.94 \pm 0.03$& 
\\
%
DBBPL & $10.2 \pm 0.6$  & $2.59 \pm 0.07$ & $2.72 \pm 0.04$ & 
1.28/717 &
$3.3 \pm 0.1$ & $1.42 \pm 0.03$ & $5.8 \pm 0.1$ & 1.55/492 \\
%
2BRMS & $6.0\, \pm\,_{0.4}^{0.6}$ & $1.27\, \pm\, ^{0.41}_{0.28}$ & 
\dots & 1.24/717 &
$3.0 \pm 0.1$ & $0.20 \pm 0.01$ & \dots & 1.51/492 \\
 & \dots & $6.81\, \pm\, ^{0.21}_{0.17} $ & \dots & & \dots & $2.78\, \pm\,_{0.09}^{0.07}$& \\   
%
2MEKAL & $10.8 \pm 0.4$ & $0.81 \pm 0.05$ & $0.45 \pm0.02$ & 1.44/721 &
$0.94\, \pm\, ^{0.04}_{0.06}$ & $0.319 \pm 0.004$ & $0.64 \pm0.04$ & 16.7/496 \\
  &\dots & $6.18 \pm 0.09$ & \dots& &\dots & $3.85\, \pm\,^{0.10}_{0.12}$ &\dots & \\ 
\noalign{\smallskip}
\hline
\end{tabular}
\end{flushleft}
\label{tab:spec_paras}
\end{table*}


\src\ was observed twice by \sax\ as a target of 
opportunity (TOO).
TOO1 took place between 1998 April 3 05:03 and 17:44 and
TOO2 between 1998 April 9 00:48 and April 10 06:49~UTC. 
The data were processed using the {\sc saxdas 1.3.0} package.  
The TOO1 exposures are 6.8, 21.5, 9.8,
and 9.3~ks in the LECS, MECS, HPGSPC, and PDS, respectively.
The corresponding count rates are 6.8, 12.7, 9.5, and 2.2 s$^{-1}$.
The TOO2 exposures are 10.7, 47.0, and 17.8~ks in the LECS, 
MECS, and HPGSPC. The corresponding count rates are
16.5, 0.8, and 0.2 s$^{-1}$. LECS and MECS data were
extracted centered on the position of \src\ using radii of 8\arcmin\ and 
4\arcmin, respectively. Background subtraction in the imaging instruments
was performed using standard files, but is not critical for such a
bright source. 
Background subtraction in the non-imaging instruments was carried out 
using data from offset intervals.

\section{The average X-ray spectra}
\label{spectrum}

Spectral analysis was performed separately on the 
average TOO1 and TOO2 NFI spectra. 
Spectra were selected in the energy ranges 0.3--10~keV, 
1.8--10~keV, 5--20~keV and 15--50~keV for the LECS, MECS, HPGSPC and PDS,
respectively.  
For TOO2 only an upper limit from the PDS in the 15--50~keV
energy range was obtained, since the source was much 
fainter $\approxgt$10~keV than in TOO1.
Factors were included in the spectral fitting to allow for known 
normalization differences between the instruments.
Uncertainties and upper limits are quoted at 90\% confidence 
throughout.
Fit results are listed in Tables \ref{tab:spec_paras} and \ref{fits}.
No simple model, e.g. absorbed power-law, broken or exponentially cut-off
power-law, thermal bremsstrahlung, or multi-temperature disk blackbody 
(\cite{mitsuda84}) plus power-law, 
gives a satisfactory fit to either observation. 
This last model has been successfully fit to the 
spectra of many soft X-ray transients (e.g., \cite{tanaka95}). 
Inspection of the residuals reveals the presence of strong emission lines
in the spectra.
Including such features in the models brings a significant, 
albeit insufficient, improvement in fit quality. 
The description of the lines is given in  Table~\ref{fits}.
All fitted lines are narrow and unresolved by the LECS and MECS. 
However a blend of narrow lines cannot be excluded. If allowed 
to vary, the Gaussian widths, $\sigma$, remain small compared to 
the instrument resolution and the fit statistics 
do not change significantly. Therefore  $\sigma$ was fixed at 0.1 keV.   
\begin{figure*}[htb]
\centerline{\psfig{figure=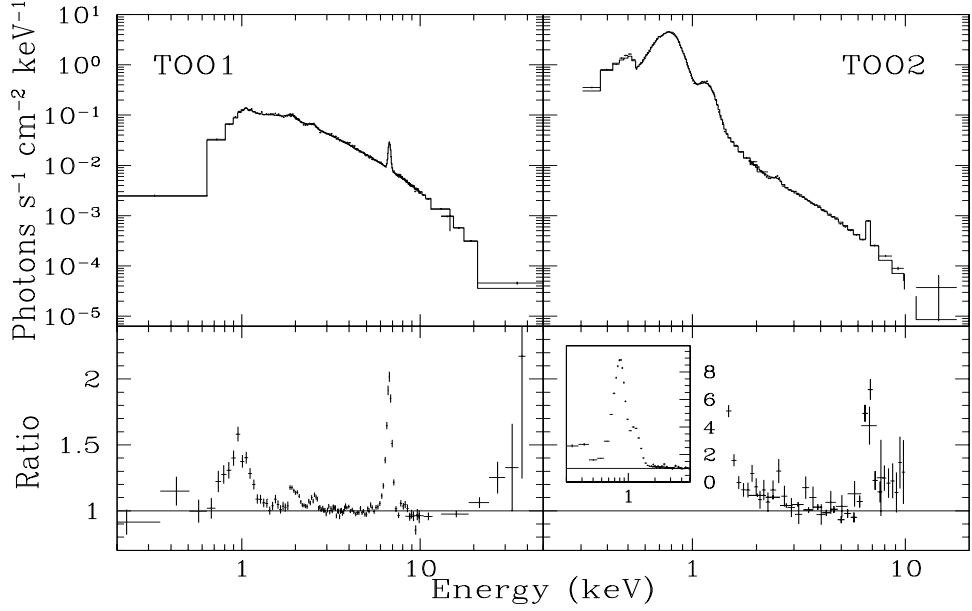,height=8.5cm,width=13.5cm,angle=-90}}
\caption{\protect \small
Top: Deconvolved photon spectra of \src\ during TOO1 and TOO2 using two 
bremsstrahlung components plus emission lines
(see Table \ref{fits}). Bottom: data to model ratios, where 
the line normalizations are set equal to zero in the models. The inset
shows the TOO2 model ratio near 1 keV where there is strong soft emission}
\label{spec_change}
\end{figure*}

\vspace{-0.2mm}

The best fit for TOO1 is obtained using a cut-off power-law with narrow
Gaussian emission features. 
The fit is formally unacceptable with a $\chi^2$ of 1.23, 
for 718 degrees of freedom (dof), but models the 
overall shape of the 1--20~keV spectrum reasonably well.  
Next best is a model consisting of two thermal bremsstrahlung
components plus  emission lines.
The observed TOO1 fluxes F$_{0.5-2}$ and F$_{2-10}$ are 0.3 and 
$1.0\times 10^{-9}$ ergs cm$^{-2}$ s$^{-1}$. At a distance of 1~kpc 
these correspond to luminosities of $3.4\times 
10^{34}$ and $1.2\times 10^{35}$~erg~s$^{-1}$. 

The X-ray spectrum of \src\ changed dramatically between TOO1 and TOO2
(see Fig.~\ref{spec_change}) with the appearance of strong soft 
emission at $\approxlt$1~keV.  
All the models listed in Table~\ref{tab:spec_paras} 
show a reduction in N$_{\rm {H}}$ of at least a 
factor $\ge$1.8 between TOO1 and TOO2.
Such a change may result from obscuration by material in an
expanding shell.
The best description of the TOO2 data is achieved with a
double bremsstrahlung model including narrow Gaussian emission lines.
The observed TOO2 fluxes F$_{0.5-2}$ and F$_{2-10}$ are 
1.7 and 0.05 $\times 10^{-9}$ ergs cm$^{-2}$ s$^{-1}$.
The double {\sc mekal} model, which fits the TOO1 spectrum 
reasonably well, provides a very poor fit to TOO2.

The double bremsstrahlung plus narrow emission lines model gives a 
reasonable and ``simple'' parameterization of both
spectra and was therefore chosen to compare TOO1 and TOO2.
Table~\ref{fits} shows that the continuum temperatures decreased
significantly between TOO1 and TOO2. 
Features at 1.9, 2.5 and 6.7 keV, identified with He-like K$\alpha$ 
emission from Si, S, and Fe, are observed in both spectra. 
There are no large changes in their equivalent widths, EW, or
mean energies between the observations. 
In TOO1 a feature is present at 0.99~keV with an EW of 163~eV.
This may be identified with K$\alpha$ emission from Ne~{\sc x} 
and/or a number of Fe-L transitions. In TOO2 intense features are present
at 0.74 and 1.15~keV with EWs of 1420 and 635~eV, respectively. 
There are no prominent emission lines with energies close to 0.74~keV.
However, if the 1.15~keV feature is interpreted as the Doppler shifted
Ne~{\sc x}/Fe-L complex (observed at 0.99~keV in TOO1), and if the
0.74~keV feature is also Doppler shifted by the same amount, its rest 
energy is 0.63~keV -- close to the
energy of the prominent O~{\sc viii} line at 0.65~keV. 
We therefore tentatively identify
the TOO2 0.74~keV feature with blue-shifted O~{\sc viii} emission. The
upper limit to a Gaussian emission feature at 0.65~keV during TOO1 is
36~eV.
Finally, the TOO1 fits improve significantly (at 99\% confidence
using the F-statistic) when a Ca {\sc xix} line is included at 3.9~keV.
However, the EW is small ($16\pm 7$ eV) and this line is not required in 
TOO2.

\begin{figure}[t]
\hbox{\psfig{figure=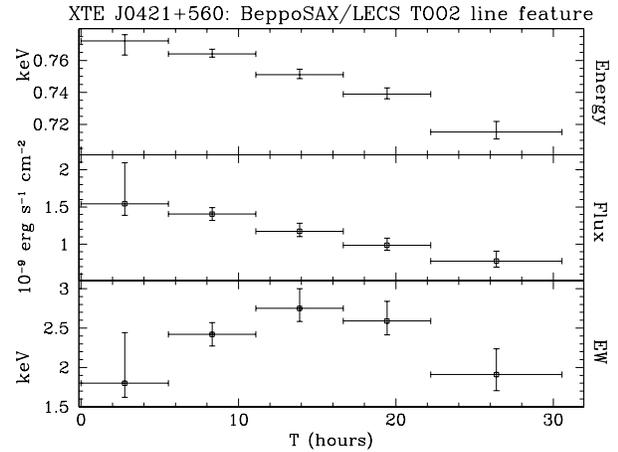,height=6.cm,width=8.cm,angle=-90} }
\caption{\protect \small
The variation in energy, flux and EW of the 0.74~keV 
feature during TOO2. The variations in these
parameters for the 1.15~keV feature are almost identical}
\label{softline}
\end{figure}

 \begin{table}
\caption[]{Two bremsstrahlung and narrow Gaussian 
emission lines model fits to \src. The O~{\sc viii} and Ne~{\sc x} features
in TOO2 are assumed to be blue-shifted from their rest energies of 0.65
and 1.02~keV (see text)}
\begin{flushleft}
\begin{tabular}{llll}
\hline\noalign{\smallskip}
Parameter                & TOO1  & TOO2 & Ident.\\ 
\noalign {\smallskip}
\hline\noalign {\smallskip}
N${\rm _H}$ (10$^{21}$ cm$^{-2})$ & $6.0 \pm\,_{0.5}^{0.6}$ & $3.0\,\pm\,_{0.4}^{0.7}$ \\
kT$_1$ (keV)             & $1.30\,\pm\,_{0.30}^{0.60}$ &    $0.20\,\pm\,_{0.02}^{0.01}$ \\
kT$_2$ (keV)             & $6.82\,\pm\,_{0.19}^{0.26}$    & $2.78\,\pm\,_{0.04}^{0.07}$ \\ \noalign {\smallskip}
\hline\noalign {\smallskip}
E$_{\rm line}$ (keV)         &0.74 &      $0.740\,\pm\,_{0.008}^{0.005}$ 
& O~{\sc viii} K$\alpha$\\
EW (eV)          & $<$45 &     $1420\,\pm\,_{240}^{525}$ &\\
\noalign {\smallskip}
\hline\noalign {\smallskip}
E$_{\rm line}$ (keV)         & $0.99\pm0.02$ &    $1.155\,\pm\,_{0.003}^{0.004}$ & Ne~{\sc x} K$\alpha$ \\
EW (eV)          & $163\,\pm\,_{36}^{55}$ &    $635\,\pm\,_{27}^{37}$ &
Fe-L\\ \noalign {\smallskip}
\hline\noalign {\smallskip}
E$_{\rm line}$ (keV)         & $1.91\pm0.02$ &    $1.86\pm0.03$ & 
Si~{\sc xiii} K$\alpha$\\
EW (eV)          & $65\pm8$ &     $105\pm 21$ &
\\ \noalign {\smallskip}
\hline\noalign {\smallskip}
E$_{\rm line}$ (keV)         & $2.50\pm0.03$ &     $2.47\pm0.03$ 
& S~{\sc xv} K$\alpha$\\
EW (eV)          & $38\pm7$ &     $89\,\pm\,_{17}^{25}$ 
\\ \noalign {\smallskip}
\hline\noalign {\smallskip}
E$_{\rm line}$ (keV)         & $6.73\pm0.01$ &     $6.75\pm0.03$ & 
Fe~{\sc xxv} K$\alpha$\\
EW (eV)          & $597\pm18$ &      $731\pm72$ &
\\ \noalign {\smallskip}
\hline\noalign {\smallskip}
$\chi_{\nu}^2$ (dof)      & 1.24 (714) &     1.53 (487) \\ 
\noalign {\smallskip}
\hline\noalign {\smallskip}
\end{tabular}
\end{flushleft}
\label{fits}
\end{table}
%

\section{Time resolved spectral analysis}
\label{var}

The data from TOO1 and TOO2 were divided into 8 and 5 time intervals, 
respectively, and fitted with the double bremsstrahlung 
and narrow emission line model (see Table~\ref{fits}).
During TOO1 the line parameters (EW and energy, if left free) do not vary 
significantly between the intervals. The upper limit to any 6.7~keV line 
shift is 1.4\%. 
The column density, ${\rm N_H}$, remains approximately constant. 
The fluxes of the different model components were calculated for each interval.
No significant variation of the fluxes is measured, except that of the hard 
bremsstrahlung component and the 6.7 keV line, 
which both decrease smoothly by a factor $\sim$1.8.

\begin{figure}[t]
\hbox{\psfig{figure=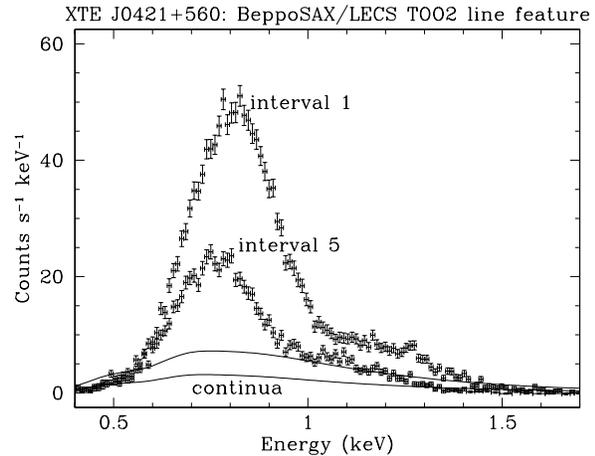,height=6.cm,width=8.cm,angle=-90} }
\caption{\protect \small
The change in the 0.74~keV feature
between TOO2 intervals 1 and 5. The two solid lines show the 
subtracted continua, changes in which cannot account for the observed shift
}
\label{lineshift}
\end{figure}

During the 30~hr observation of TOO2 the energies of the  
0.74 and 1.15~keV lines, tentatively identified as blue-shifted 
O~{\sc viii} and  Ne~{\sc x}/Fe L emission, respectively, 
both display a regular and significant 
decrease by 7.7$\pm^{1.1}_{2.1}$\% and 9.6$\pm^{1.3}_{2.4}$\%.  
Changes in the underlying continuum are unlikely to be responsible 
for these shifts but cannot be excluded.
There is no significant change in ${\rm N_H}$, 
or the energies, fluxes, and 
EWs of the 1.9, 2.5 and 6.75~keV lines.
The upper limit to any shift in the 6.75~keV line is 3.5\%.
The fluxes in the soft and hard bremsstrahlung components 
decrease by factors 1.3 and 3, and the fluxes in the
two soft lines decrease by a factor 1.8. 
The EWs of the two soft lines increase 
then decrease (see Fig.~\ref{softline}). 
Fig.~\ref{lineshift} shows the change in energy of the 0.74~keV 
line during TOO2.

\section {Discussion}
\label{subsec:discussion}

The X-ray spectra of \src\ presented here are unlike those of any other
X-ray transient. Certain properties are reminiscent of 
the galactic radio jet source SS\,433
(e.g. the appearance of a twisted radio-jet after the outburst, the 
presence of shifting emission lines and the X-ray luminosities;
see also \cite{frontera98}). 
The X-ray spectrum of SS\,443 displays a pattern of red- and
blue-shifted He- and H-like emission lines superposed on a 
two temperature bremsstrahlung continuum 
(\cite{kotani94}).
The lines most probably originate in 
two collimated precessing (with a period of 163~d) 
relativistic (0.26~$c$) jets which result from
super-Eddington accretion onto a black hole (e.g., \cite{rose95}). 
We propose that the time variable features in \src\ also originate in 
jets and that the variation in energy of these features is due to 
precession. Fitting a sinusoid to the variation in energy of the
0.74~keV feature during TOO2 and assuming a rest energy of 0.65~keV
(ie. that the line is O~{\sc viii} K$\alpha$) implies
a precessional period of $\ge$6~days and a 
velocity of $0.20\pm\,_{0.01}^{0.08}\;c$ (1$\sigma$ uncertainty).  

\asca\ observations of some high mass X-ray binary pulsars such as Vela\,X-1
(\cite{nagase94}) reveal spectra rich in He-like emission lines,
almost certainly due to reprocessing in circumstellar material. 
The stationary lines (Si~{\sc xiii}, S~{\sc xv}, Fe~{\sc xxv},
and possibly Ca~{\sc xix}) 
most probably originate in circumstellar matter. 
The identification of the moving features with emission from
H-like ions (although the rest energies are uncertain), and the 
stationary ones with He-like ions, is consistent with this interpretation.
In the case of SS\,443 two sets of moving X-ray lines are seen 
(\cite{kotani94}),
although earlier studies had revealed only one (e.g, \cite{watson86}).
This was explained by assuming that one beam was occulted by the
accretion disk at certain precession phases (\cite{stewart87}).
This may also be the case in \src.

\begin{acknowledgements}
AO acknowledges an ESA Fellowship.
The \sax~satellite is a joint Italian and Dutch programme.
\end{acknowledgements}

\end{document}